\documentclass[aps,prl,twocolumn,showpacs]{revtex4}
\usepackage{graphicx}

\begin{document}

\title{Specific-Heat Measurement of Residual Superconductivity  in the Normal State of Underdoped Cuprate Superconductors}

\author{{Hai-Hu Wen}\email{hhwen@aphy.iphy.ac.cn}, {Gang Mu, Huiqian Luo, Huan Yang,  Lei Shan, Cong Ren, Peng Cheng, Jing Yan, Lei Fang}}

\affiliation{National Laborotary for Superconductivity, Institute of
Physics and Beijing National Laboratory for Condensed Matter
Physics, Chinese Academy of Sciences, P.~O.~Box 603, Beijing 100080,
P.~R.~China}

\date{\today}

\begin{abstract}

We have measured the magnetic field and temperature dependence of
specific heat on $Bi_2Sr_{2-x}La_xCuO_{6+\delta}$ single crystals in
wide doping and temperature regions. The superconductivity related
specific heat coefficient $\gamma_{sc}$ and entropy $S_{sc}$ are
determined. It is found that $\gamma_{sc}$ has a hump-like anomaly
at $T_c$ and behaves as a long tail which persists far into the
normal state for the underdoped samples, but for the heavily
overdoped samples the anomaly ends sharply just near $T_c$.
Interestingly, we found that the entropy associated with
superconductivity is roughly conserved when and only the long tail
part in the normal state is taken into account for the underdoped
samples, indicating the residual superconductivity above T$_c$.

\end{abstract}


\maketitle

One of the most important issues in cuprate superconductors is the
existence of a pseudogap above $T_c$ in the underdoped
region.\cite{1} It appears in close relationship with many anomalous
properties in the normal state, and thus receives heavy debate about
its nature. One scenario assumes that the pseudogap reflects only a
competing or coexisting order of superconductivity and it may have
nothing to do with the pairing. However, other pictures, such as the
Anderson's resonating-valence-bond (RVB) model\cite{2} and related
models\cite{3,4} regard the pseudogap as due to the spin-singlet
pairing in the spin liquid state and it has a close relationship
with Cooper pairing for superconductivity. Experimentally some
evidence for fluctuating superconductivity in the normal state of
underdoped samples have been inferred in the measurements of Nernst
effect,\cite{5,XuZA} diamagnetization,\cite{6} time-domain optical
conductivity\cite{7} and thermal expansion,\cite{8} etc. \emph{The
evidence from specific heat (or entropy) for this residual
superconductivity in the normal state is, however, still lacking.}

By using the differential heat capacity technique, Loram et
al.\cite{9} successfully measured the electronic specific heat (SH)
of cuprate superconductors (most of time at zero field). The
advantage of this technique made it possible to observe the SH
anomaly near $T_c$ and the suppression to the electronic SH
coefficient $\gamma_e$ below $T^*$ in underdoped region. It remains,
however, unresolved whether this suppression to $\gamma_e$ below
$T^*$ is due to the preformed pairing, or induced solely by the
competing order.\cite{10} In addition, for a superconductor within
the BCS scenario, the superconductivity related entropy (SRE) is
conserved at just above $T_c$. It is thus also curious to know
whether the SRE is conserved in very underdoped samples. Answering
this question casts big challenge since the SRE is difficult to be
determined in cuprate superconductors. One way to reach this goal is
to measure the difference of heat capacity between the
superconducting state and a normal state background which is
normally achieved by using a high magnetic field to suppress the
superconductivity. The heat capacity under magnetic fields has been
measured near $T_c$ by Junod et al. on YBCO, Bi-2212 and Bi-2223
single crystals.\cite{11} Due to the very high critical field in
those samples, the relatively low magnetic field (about 10 Tesla) in
the usual laboratory cannot suppress the bulk superconductivity
completely. It is thus highly desired to do the field dependent SH
measurement on some single crystals with low $T_c$, in such case a
magnetic field in the scale of 10 Tesla can suppress the bulk
superconductivity. As far as we know, no such investigations on SH
on systematic doped cuprate samples have been reported. In this
Letter, we present the SH data measured on high quality
$Bi_2Sr_{2-x}La_xCuO_6$ (Bi-2201) single crystals\cite{12} in wide
temperature and doping regime and the superconductivity is tuned by
the magnetic field. \emph{The evidence
 for residual superconductivity far above $T_c$ has been
found based on the analysis of entropy conservation in underdoped
samples.}

\begin{figure}
\includegraphics[width=8cm]{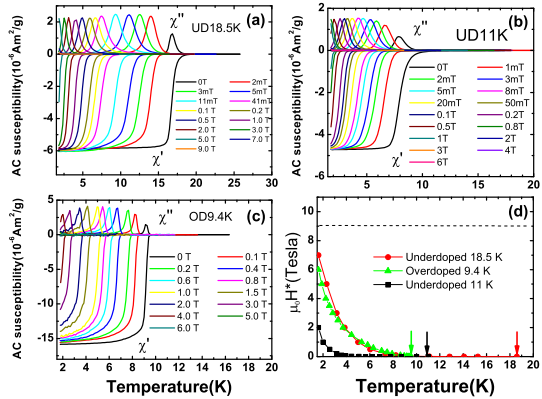}
\caption{(Color online) AC susceptibility for three single crystals
of (a) UD18.5K, (b) UD11K and (c) OD9.4K. The measurements were done
with an AC field of 0.1 Oe, and oscillating frequency of 333 Hz. The
critical field H* for bulk superconductivity (see text) is shown in
(d). The arrows indicate the positions of the bulk superconducting
transitions at zero field for the three samples. In this study all
measurements were done with the magnetic field parallel to c-axis of
the crystals.
 } \label{fig1}
\end{figure}

In this experiment we have selected six high quality crystals grown
by the traveling solvent floating zone technique,\cite{12} five of
them are from $Bi_2Sr_{2-x}La_xCuO_{6+\delta}$ with x = 0.8
(underdoped, p $\approx$ 0.11, $T_c$ = 11 K), x = 0.7 (underdoped, p
$\approx$
 0.123, $T_c$ = 18.5 K), x = 0.6 (underdoped, p $\approx$0.131, $T_c$ = 22 K),
x = 0.4 (optimally doped, p $\approx$ 0.16, $T_c$ = 30 K), x = 0.1
(overdoped, p $\approx$ 0.20, $T_c$ = 17.6 K), and one of
$Bi_{1.74}Sr_{1.88}Pb_{0.38}CuO_{6+\delta}$ (overdoped, p$\approx$
0.22, $T_c$ = 9.4 K). For simplicity, they are denoted as UD11K,
UD18.5K, UD22K, OP30K, OD17.6K and OD9.4K, respectively. In Fig.1 we
present the AC susceptibility of two underdoped samples in (a) and
(b), one overdoped sample (with Pb doping) in (c).  For the
underdoped samples, see for example Fig.1(b), a very small magnetic
field can suppress the superconducting transition quickly
manifesting a very fragile superfluid density. If we take the point
where both the real part susceptibility $\chi$' and the imaginary
part $\chi$'' merge into the flattened normal state background
(actually buried in the noise level) as the criterion for bulk
superconductivity, the critical field H*(T) is obtained and shown in
Fig.1(d). One can see that, when the field is beyond 9 Tesla, no
bulk superconductivity can be detected above 2 K. This allows to use
the data at 9 Tesla as the appropriate background for the state
without bulk superconductivity above 2 K.\cite{Background} Thus we
define the superconductivity related SH as
$\gamma_{sc}=[C(H)-C(9T)]/T$, here C(H) and C(9T) are the total heat
capacity measured at the magnetic field H and 9 T, respectively.
This treatment naturally removes the phonon contribution since it is
field independent.

\begin{figure}
\includegraphics[width=9cm]{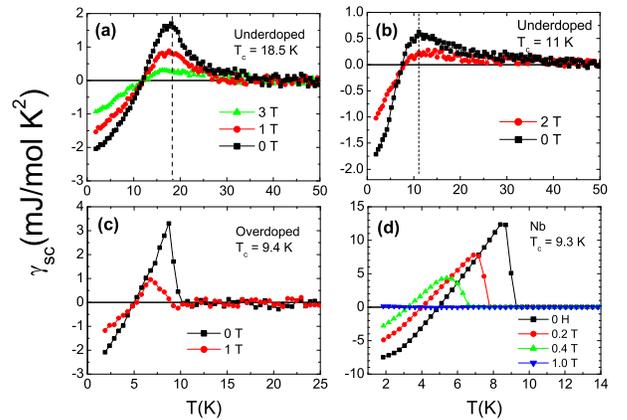}
\caption{(Color online) The subtracted specific heat  for four
samples: (a) UD18.5K, (b) UD11K, (c) OD9.4K and (d) $Nb$ with $T_c$
= 9.3 K (using 2 T as the background). In (a) and (b) the dashed
lines mark the positions of $T_c$. } \label{fig2}
\end{figure}

Fig.2 presents the temperature dependence of $\gamma_{sc}$ for the
corresponding samples shown in Fig.1. The heat capacity was measured
by using the relaxation method based on a PPMS system (Quantum
Design) with the latest upgraded puck. For the underdoped samples,
one can easily draw the following interesting conclusions: (1) In
the zero temperature approach, the magnetic field always enhances
$\gamma_{sc}$, leading to a finite quasiparticle density of states
(DOS). This is consistent with the results in $La_{2-x}Sr_xCuO_4$
and other systems.\cite{15,16} Our results support also the
conclusion of a Fermi surface in the normal state that revealed by
recent quantum oscillation measurements.\cite{17} (2) What surprises
us is that there is NO step-like SH anomaly at $T_c$ for the
underdoped samples, instead it shows a broad hump-like peak at about
$T_c$ and remains as a long tail of $\gamma_{sc}(T)$ far above
$T_c$. For example, for the underdoped sample with $T_c$ = 11 K,
this long tail can last up to about 42 $\pm$ 5K where the signal is
buried in the noise background. (3) In a BCS superconductor, when
the superconductivity is suppressed by a magnetic field, the peak
height of the SH anomaly is suppressed and the transition
temperature is lowered due to the field induced pair-breaking (see
an example in Fig.2(d) for a conventional BCS superconductor Nb).
However, as shown in Fig.2(a) and (b), for the underdoped samples,
one can see that the position of the SH peak keeps unchanged but the
height is suppressed greatly by the magnetic field. Very
surprisingly, the onset for bulk superconductivity as measured by
the ac susceptibility shifts quickly with the magnetic field. This
indicates that the bulk superconductivity is not determined by the
position of the SH anomaly. Regarding the long tail of
$\gamma_{sc}(T)$ extending up to high temperatures, we conclude that
there is residual superconductivity far above $T_c$. In order to
check whether this is a special property for the underdoped samples,
in Fig.2(c), we present the data for a heavily overdoped sample in
the same system. It is easy to see that the $\gamma_{sc}(T)$ data
shows only a step-like BCS mean field transition with the absence of
the long tail in the normal state.

To further illustrate the difference between the underdoped and
overdoped samples, we present the  $\gamma_{sc}(T)$ data in Fig.3(a)
and Fig.3(b). For underdoped samples, the long tail of
$\gamma_{sc}(T)$ extends to the temperature region between 35 K and
45 K. In addition, towards underdoping, the SH peak is strongly
suppressed leading to a hump-like anomaly. For the strongly
underdoped sample, UD11K, the ratio of $\Delta C/\gamma_nT_c$ = 0.25
determined here is far below the value expected by the BCS theory (
$\Delta C /\gamma_nT_c$ = 1.43 for an s-wave gap and higher for a
d-wave gap), where we take -$\gamma_{sc}(0)$ as $\gamma_n(0)$ and
$\Delta C = \gamma_{sc}(T_c)T_c$. When the hole concentration
increases, the ratio is getting larger, but for all underdoped
samples, this ratio is significantly below the expected BCS value.
Since the applied magnetic field is not high enough to suppress the
bulk superconductivity for the optimally doped sample, the data were
shown only above 15 K, and the $\gamma_{sc}(T)$ tail extends to
about 42 K which is close to the upper boundary of the Nernst signal
in this sample.\cite{5} It is interesting to note that the SH
anomaly near $T_c$ is not sharp-step like for the optimally doped
sample, rather it shows a symmetric peak. This is consistent with
the observation by Junod et al. in optimally doped Bi-2212 and
Bi-2223.\cite{18} For overdoped samples, this tail becomes much
shorter: for sample OD17.6 K, it ends at about 23 K, and for the
very overdoped one OD9.4K, it vanishes at 10 K being very close to
$T_c$ = 9.4 K. In Fig.3(c) we present the temperature dependence of
the entropy calculated by $S_{sc}=\int_{0}^{T}\gamma_{sc}(T')dT'$,
here the data of $\gamma_{sc}(T)$ at T = 0 K was obtained by doing
the linear extrapolation of the low temperature data. For the
overdoped sample OD9.4K, the entropy is conserved at just $T_c$ =
9.4 K. The slight nonzero entropy above $T_c$ is induced by the
uncertainty in deriving the value of $\gamma_{sc}(T)$ at T = 0 K.
The condensation energy calculated by integrating the entropy,
i.e.,$E_{cond}=-\int_{0}^{T_c}S_{sc}(T')dT'$, is about 48 $\pm$ 5
mJ/mol for the sample OD9.4K. For the underdoped sample UD18.5K, the
entropy is obviously not conserved by integrating  $\gamma_{sc}(T)$
just up to $T_c$, but surprisingly, it becomes roughly conserved
when the long tail part of $\gamma_{sc}(T)$ in the normal state is
taken into account as shown by the red circles in Fig.3(c). As
stressed previously,\cite{19,20} in underdoped cuprates, the term
"condensation energy" may have a different meaning as compared to a
conventional superconductor since the pairing in the normal state
certainly contributes a significant part to the total condensation
energy, but the bulk superconducting transition at $T_c$ saves extra
energy. By integrating the entropy from T to 50 K, namely
$E_{cond}=-\int_{T}^{50K}S_{sc}(T')dT'$, we derived the temperature
dependence of the condensation energy $E_{cond}$ for three
underdoped samples UD11K, UD18.5K and UD22K, and the heavily
overdoped sample OD9.4K (integral from T to 18 K). The results are
shown in Fig.3(d). For the sample UD18.5K the total condensation
energy at T = 0 K is about 263 $\pm$ 10 mJ/mol, while the normal
state contributes an energy-saving of about 52 $\pm$ 5 mJ/mol, this
gives a portion about 20\% of the total "condensation energy". An
estimate for the more underdoped sample UD11K finds that the normal
state contribution to the total "condensation energy" can be as
large as 54\%, as shown by the blue triangles in Fig.3(d). This
large ratio of the normal state contribution to the condensation
energy makes it almost impossible to attribute the residual
superconductivity above $T_c$ to the Gaussian fluctuation. It also
clearly prohibits from understanding the superconducting transition
in underdoped samples within the BCS scenario.

\begin{figure}
\includegraphics[width=8cm]{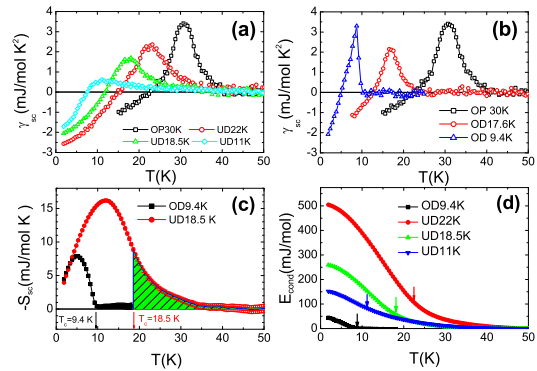}
\caption{(Color online) A collection of  $\gamma_{sc}(T)$ at zero
field for three underdoped samples and one optimally doped sample in
(a), two overdoped samples and one optimally doped sample in (b).
(c) Temperature dependence of the superconductivity related entropy
calculated by integrating $\gamma_{sc}(T)$ in wide temperature
region. (d) The condensation energy calculated through integrating
the entropy (see text). The arrows mark the temperatures of the bulk
superconducting transition. } \label{fig3}
\end{figure}

\begin{figure}
\includegraphics[width=8cm]{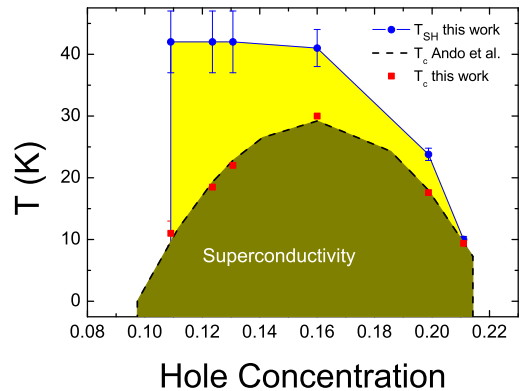}
\caption{(Color online) A generic phase diagram plotted based on the
specific heat data. The dashed line is the $T_c$-p plot from Ando's
group in the same system. The red squares represent the measured
$T_c$ values of our samples at the same nominal doping level. The
blue circles show the temperatures $T_{SH}$ where $\gamma_{sc}$(T)=
0 $\pm$ 0.15 $mJ/molK^2$ (within the error bars of the experiment).
One can see that the gap between $T_c$ and $T_{SH}$ is getting
monotonically larger but $T_{SH}$ flattens out in more underdoped
region. } \label{fig4}
\end{figure}

In Fig.4, we present a generic phase diagram derived from our data.
 Here we used the empirical relation p=0.21-0.18x to obtain
the hole concentration.\cite{21} The red squares represent the $T_c$
values of our samples, which show very good consistency with that of
Ando et al..\cite{21} The blue circles show the vanishing points
$T_{SH}$ of the long tail of $\gamma_{sc}(T)$ using the criterion of
0$\pm0.15$ $mJ/molK^2$, where the SRE has dropped below 0.5
$mJ/molK$ (see Fig.3(c)). One can see that the difference between
$T_c$ and $T_{SH}$ is getting monotonically larger towards
underdoping. This phase diagram looks qualitatively similar to that
depicted based on the Nernst measurements,\cite{5,22} but the upper
limit temperatures for the Nernst signal on underdoped samples are
higher than the values derived from our specific heat. There is a
possible explanation about this discrepancy: It was argued by the
Princeton group that the normal state Nernst signal comprises both
the coherent part and incoherent part.\cite{22} The upper boundary
of temperature for the coherent part is much lower than the
incoherent one. Our data $\gamma_{sc}(T)$ here measures the residual
superconductivity, thus correspond well with the coherent part of
the Nernst signal. Since the entropy is naturally conserved if the
normal state part of $\gamma_{sc}(T)$ is taken into account, we thus
believe that there is residual superconductivity in the normal state
of underdoped samples.

Our results are also qualitatively consistent with the recent
observation of local pairing above $T_c$ as seen by the scanning
tunneling microscopy.\cite{23} These nano-scale droplet of Cooper
pairs above $T_c$ will certainly contribute to the condensation
energy of the system and make the entropy unconserved (at $T_c$)
unless the upper temperature for counting the entropy is beyond
$T_{SH}$ in our definition. In this sense the superconducting
transition in underdoped samples means to establish the long range
phase coherence.\cite{3} Thus the thermal energy $k_BT_c$ may be
equated by the phase coherence energy
$E_{coh}=\hbar^2\rho_s(T_c)/$m* given by Deutscher et al.,\cite{24}
where $\rho_s$ is the superfluid density, m* is the effective mass.
Below $T_c$ the quasi-particles which reside on the small Fermi
surfaces in the normal state\cite{17,25} will pair up each other and
condense into the superconducting state together with the residual
Cooper pairs formed above $T_c$. This naturally builds up a new gap
on the small Fermi surfaces in the region near the
nodes.\cite{26,27} Above $T_c$, strong phase fluctuation\cite{3,28}
breaks up many Cooper pairs with small pairing energy,\cite{25} but
some residual pairs with stronger pairing strength will exist up to
a high temperature. As demonstrated by our data, the superconducting
condensation in the underdoped region cannot be put into the BCS
category.

In summary, the specific heat anomaly at $T_c$ is strongly
suppressed through underdoping leading to a hump-like anomaly with
the height much below the value predicted by the BCS theory. A long
tail of $\gamma_{sc}(T)$ has been found far into the normal state
for underdoped samples. The entropy calculated by integrating
$\gamma_{sc}(T)$ to $T_c$ is clearly not conserved, but it becomes
roughly conserved when and only the tail part in the normal state is
taken into account. These results prohibit from using the BCS
picture to understand the superconducting transitions in underdoped
samples.

We thank J. Zaanen, J. Tallon and P. W. Anderson for comments and
suggestions. We acknowledge also S. Kivelson, F. C. Zhang, Z. Y.
Weng, Q. H. Wang, G. Appeli, P. C. Dai and Y. Y. Wang for useful
discussions. We thank L. Zhao, G. D. Liu and X. J. Zhou for
providing us one as-grown sample (OD9.4K). This work was supported
by the MOST of China (973 Projects No.2006CB601000, No.
2006CB921802) and CAS Project.


\begin{thebibliography}{00}
\bibitem{Background}To define a "normal state background" is a non-trivial
issue for underdoped cuprate superconductors. Here we use the data
measured at 9 T as a relative but appropriate background because the
AC susceptibility shows no trace of bulk superconductivity above 2
K.  Furthermore, we found that the difference between the specific
heat data of 9T and 8T is almost invisible.
\end{thebibliography}
\end{document}